\documentclass[nologo,url,11pt,a4paper]{ETHpaper}
\usepackage{graphics}
\usepackage{amssymb}

\begin{document}
\title{Global firing induced by network disorder in ensembles of active
rotators}
\author{Claudio J. Tessone$^1$, Dami\'an H. Zanette$^2$, and Ra\'ul
Toral$^3$}
\address{
$^1$Chair of Systems Design, ETH Zurich, Kreuzplatz 5, CH-8032 Zurich,
Switzerland \\
$^2$Consejo Nacional de Investigaciones Cient\'{\i}ficas
y T\'ecnicas, Centro At\'omico Bariloche and Instituto Balseiro,
8400 Bariloche, R\'{\i}o Negro, Argentina \\
$^3$IFISC (Instituto de F{\'\i}sica Interdisciplinar y Sistemas Complejos),
CSIC-UIB, Ed. Mateu Orfila, Campus UIB, E-07122 Palma de Mallorca, Spain }
\date{\today}
\authoralternative{Claudio J. Tessone, Dami\'an H. Zanette, and Ra\'ul
Toral}


\reference{\emph{Submitted}. Cite as: \textsl{arXiv:0710.3311v1}}


\www{\url{http://www.sg.ethz.ch}}

\maketitle


\abstract{We study the influence of repulsive interactions on an
ensemble of coupled excitable rotators. We find that a moderate
fraction of repulsive interactions can trigger global firing of the
ensemble. The regime of global firing, however, is suppressed in
sufficiently large systems if the network of repulsive interactions
is fully random, due to self-averaging in its degree distribution.
We thus introduce a model of partially random networks with a broad
degree distribution, where self-averaging due to size growth is
absent. In this case, the regime of global firing persists for large
sizes. Our results extend previous work on the constructive effects
of diversity in the collective dynamics of complex systems.

\section{Introduction}

Noise and disorder can have a variety of effects on the collective
dynamics of complex systems such as biological populations, chemical
reactions, oscillator ensembles, among others. Somehow
paradoxically, such effects often play a constructive role in
inducing coherent behaviour in the system \cite{GS,SSG,TTL}. An
already classical example is the phenomenon of stochastic resonance,
where noise of appropriate intensity enhances the response of a
system to an oscillatory external forcing \cite{stochres}. A similar
effect in an extended system can be caused by the presence of some
degree of heterogeneity amongst the constituents, a phenomenon named
{\sl diversity-induced resonance} \cite{TTMG}. Also, a certain
degree of structural disorder in a network can make the propagation
of information considerably more efficient \cite{rum}.

It was recently shown that, in an ensemble of coupled excitable
rotators, noise and disorder are able to trigger global firing
\cite{TSCT:2007}. During these events, a substantial fraction of the
ensemble is excited almost synchronously, so that the joint firing
of all those elements gives rise to a collective process which can
be detected at macroscopic level. This phenomenon has been observed
under the action of additive noise, when stochastic terms are added
to the evolution equations, or under ensemble heterogeneity, when
the parameters that define the individual dynamics of the rotators
differ all over the ensemble. Since these
previous studies considered all-to-all (global) homogeneous coupling
between elements, the effect of disorder in their interaction
network was not analysed.

Global coupling has frequently been invoked as an analytically
tractable interaction pattern for which several systems are
completely solvable \cite{kuramoto}. In real situations, however,
the interaction pattern is usually heterogeneous, with different
weights for each pair of interacting elements. Some systems, such as
neural networks, even require the consideration of interactions of
different signs, to account for excitatory and inhibitory coupling
\cite{KS:1998,DA:2001,TUZKT}. Interaction patterns with attractive
and repulsive weights in ensembles of coupled dynamical elements was
first addressed by Daido \cite{HD:1987,HD:1992,HD:2000}. Their
effect on the stability of synchronisation in an ensemble of phase
oscillators, in terms of the relative number and strength of
positive and negative couplings, has been studied recently
\cite{DHZ:2005}. It was found that a certain fraction of
sufficiently strong repulsive interactions, distributed at random
over the interaction pattern, is able to induce a transition from
the highly coherent state of full synchronisation, where all
elements oscillate in phase, to an unsynchronised state.

The presence of randomly scattered negative couplings in the
interaction pattern of an ensemble of dynamical elements is a source
of structural disorder, complementary to noise and to the
heterogeneity of the ensemble. It is therefore natural to study its
effects on the collective dynamics of coupled excitable elements
and, in particular, on the phenomenon of global firing. This is our
aim in this paper, which is organised as follows: In section
\ref{sect:2}, we introduce the model of coupled excitable rotators,
and the order parameters used in the characterisation of the
collective macroscopic behaviour of the ensemble. Next, in section
\ref{sect:erdos}, we consider the case when repulsive interactions
form a fully random (Erd\H{o}s-R\'enyi) network. In this case,
global firing is observed for finite-size systems but, due to
self-averaging in the degree distribution of fully random networks,
the phenomenon is suppressed in the limit of infinitely large
ensembles. Thus, in section \ref{sect:hier}, we introduce a model of
partially random networks where the relative dispersion of the
degree distribution does not depend on the size. When repulsive
interactions are distributed over such a network, global firing is
found to persist for large systems. Our conclusions are discussed in
the final section \ref{sect:conc}.

\section{Model and order parameters} \label{sect:2}

We consider an ensemble of coupled active rotators \cite{SK:1986}
with individual phases $\phi_j(t) \in [0,2\pi)$, $j=1,\dots,N$,
whose dynamics is given by
\begin{equation}\label{eq:ard:phidot}
\dot{\phi_j} = \omega -  \sin \phi_j + \frac{C}{N} \sum_{k=1}^N
W_{kj} \sin (\phi_k-\phi_j ).
\end{equation}
The coupling strength is measured by the parameter $C\ge 0$, and
each factor $W_{kj}$ weights the interaction of a specific rotator
pair. These weights are symmetric, $W_{kj}=W_{jk}$. Attractive and
repulsive interactions are characterised, respectively, by
$W_{kj}>0$ and $W_{kj}<0$.

The natural frequency $\omega>0$ is the same for all rotators. For
$\omega > 1$ and in the absence of coupling, the individual dynamics
 is oscillatory, with an actual frequency $\omega' =
\sqrt{\omega^2-1}$. The case $\omega<1$ corresponds to
excitable individual dynamics: for $C=0$, the phase $\phi_j$ of each
rotator has two fixed points, one of them stable ($\phi_s <\pi/2$)
and the other unstable ($\phi_u >\pi/2$), at the two solutions of
$\sin \phi_j = \omega$. A perturbation of the stationary stable
solution which overcomes the unstable fixed point $\phi_u$ gives
rise to the {\it firing} of the rotator, in the form of a long phase
excursion which finally returns to the rest state $\phi_s$.
Throughout this paper we focus the attention on the excitable regime
$\omega<1$.

In our model noise is absent and, for $C=0$, individual rotators are
identical. Diversity is thus restricted to disorder in the
interaction network, through the weights $W_{kj}$. Our main aim is,
in fact, to analyse the effects of this source of diversity in the
collective dynamics of the system.

If all the phases $\phi_j$ coincide with one of the two fixed points
of the individual dynamics, either $\phi_s$ or $\phi_u$, the
ensemble is in a stationary state --which we refer to as {\it full
synchronisation}-- for any value of $C$. If the weights $W_{kj}$ are
positive for all $k$ and $j$, the fully synchronised state where all
the rotators are in the rest state, $\phi_j=\phi_s$ for all $j$, is
stable. The stability of this collective rest state can break down,
however, in the presence of repulsive interactions, when some of the
weights $W_{kj}$ are negative and their absolute value is large
enough. In this situation, generally, the ensemble does not reach an
asymptotic stationary state. Individual phases can now rotate
irregularly around their whole domain and, as we show below, a
regime of global firing --where a fraction of the ensemble is
collectively entrained into long excursions from the unstable fixed
point to the rest state-- becomes possible.

The collective behaviour of the rotator ensemble, including possible
transitions between different dynamical regimes, is well
characterised by a set of order parameters defined in terms of the
individual phases $\phi_j(t)$. First, we take the average of the
imaginary phase exponentials
\begin{equation}\label{eq:rho}
\rho(t) \exp [ i \Psi(t)] = \frac{1}{N}\sum_{j=1}^N \exp [i \phi_j
(t)] ,
\end{equation}
and compute the Kuramoto order parameter as $\rho \equiv \langle
\rho(t) \rangle$, where $\langle \cdot \rangle$ stands for the time
average over a long interval  \cite{kuramoto}. This parameter is a
direct measure of the degree of synchronisation attained by the
ensemble. For a fully synchronised state we have $\rho=1$, whereas
$\rho \sim 1/\sqrt N$ for a state where phases are uniformly
distributed over $[0,2\pi)$.

The Kuramoto order parameter cannot discern between the case where
phases are synchronised at a fixed point, as at the rest state
$\phi_s$, and the case where they rotate coherently, as expected to
occur in a regime of global firing. To discriminate between static
and dynamic entrainment, we apply the order parameter introduced by
Shinomoto and Kuramoto \cite{SK:1986}:
\begin{equation}\label{eq:zeta}
\zeta = \left< \left| \rho(t) \exp [i\Psi(t)] - \left< \rho(t)\,
\exp [i\Psi(t)] \right> \right|  \right>.
\end{equation}
This parameter differs from zero for synchronous firing only.

A third order parameter, frequently used in the analysis of
stochastic transport\cite{R02}, is the current
\begin{equation}\label{eq:j}
J=\frac{1}{N}\sum_{j=1}^N\left < \dot\phi_j(t)\right>.
\end{equation}
It measures the level of (not necessarily synchronised) firing over
the whole ensemble.

\section{Erd\H{o}s-R\'enyi networks of repulsive links}
\label{sect:erdos}

First, we consider an ensemble of rotators governed by Eq.
(\ref{eq:ard:phidot}), whose interaction weights are distributed at
random according to the following prescription:
\begin{equation} \label{eq:wkj}
W_{kj} = \left\{
\begin{array}{rl}
1 & \hbox{with probability } 1-p_d , \\
- \kappa & \hbox{with probability }  p_d.
\end{array}
\right.
\end{equation}
The coefficient $\kappa>0$ measures the relative strength of
repulsive and attractive interactions, and the probability $p_d$
fixes the expected fraction of negative weights. Repulsive
interactions define a network which, with the prescription of Eq.
(\ref{eq:wkj}), has the same structure as an Erd\H{o}s-R\'enyi
random network \cite{ER:1959,ER:1960,AB:2002}. On the average, the
network of repulsive interactions has $p_d N (N-1)/2$ links. We call
${\cal N}_j$ the set of neighbours of rotator $j$ in that network.
The mean number of rotators in ${\cal N}_j$ is $p_d (N-1)$, and the
expected dispersion around this average is of order $\sqrt{N}$, so
that the relative dispersion decreases with the size as
$1/\sqrt{N}$.

It is useful to rewrite Eq. (\ref{eq:ard:phidot}) in terms of the
quantities $\rho(t)$ and $\Psi(t)$ introduced in Eq. (\ref{eq:rho}),
as
\begin{eqnarray}
\dot{\phi_j} = \omega &-&  \sin \phi_j + C \rho\, \sin(\Psi-\phi_j)
\nonumber \\ &-& \frac{C(1+\kappa)}{N} \sum_{k \in \mathcal{N}_j}
\sin\left(\phi_k-\phi_j \right). \label{ard:phidotredux}
\end{eqnarray}
This equation describes the evolution of $\phi_j(t)$ as governed by
the interaction with a single (mean-field) oscillator with phase
$\Psi (t)$ with an effective coupling strength $C\rho (t)$ plus a
negative contribution with coupling $C(1+\kappa)$ from the
neighbourhood of rotator $j$ in the network of repulsive links.

\subsection{Stability of the fully synchronised state}
\label{sect:synchr}

As advanced above, in the presence of repulsive interactions, the
state of full synchronisation may be unstable. The stability
condition can be obtained from linearisation of Eq.
(\ref{ard:phidotredux}). Writing $\phi_j(t) = \phi_s+\delta_j (t)$,
we get for the deviations from the fixed point the equations
\begin{equation}\label{eq:delta:ardis}
\dot{ \delta_j }= - \sqrt{1-\omega^2}\, \delta_j - C \delta_j -
\frac{C (\kappa+1) }{N} \sum_{k \in \mathcal{N}_j} ( \delta_k -
\delta_j ).
\end{equation}
Full synchronisation is stable if all the eigenvalues of the matrix
\begin{equation} \label{eq:matj}
 \mathbb{J} = - ( \sqrt{1-\omega^2} + C )
\mathbb{I} - \frac{C (\kappa+1) }{ N}  ( \mathbb{M} - \mathbb{N} )
\end{equation}
are negative or have negative real parts. Here, $\mathbb{I}$ is the
identity matrix, $\mathbb{M}= \{ M_{kj} \}$ is the adjacency matrix
of the network of repulsive links (defined as $M_{kj}=1$ if there is
a repulsive link between rotators $k$ and $j$, and $M_{kj}=0$
otherwise), and $\mathbb{N} = \{ N_{kj} \}$ is a diagonal matrix,
where $N_{jj}\equiv k_j = \sum_i M_{ij}$ is the number of rotators
in ${\cal N}_j$ --namely, the degree of site $j$.

Whether the stability condition for full synchronisation is
fulfilled or not depends on the specific realisation of the network
of repulsive links. Even for the same probability $p_d$, and for
fixed values of $C$, $\omega$ and $\kappa$, the stability of full
synchronisation depends on the particular distribution of negative
weights. The uppermost panel of Fig. \ref{fig1} shows numerical
results for the fraction $f_d$ of realisations of the network of
repulsive links for which full synchronisation is unstable, as a
function of $p_d$. As expected, $f_d$ grows as the number of
repulsive links increases. Different curves correspond to different
system sizes $N$.

\begin{figure}\begin{center}
\resizebox{.37\columnwidth}{!} {\includegraphics*{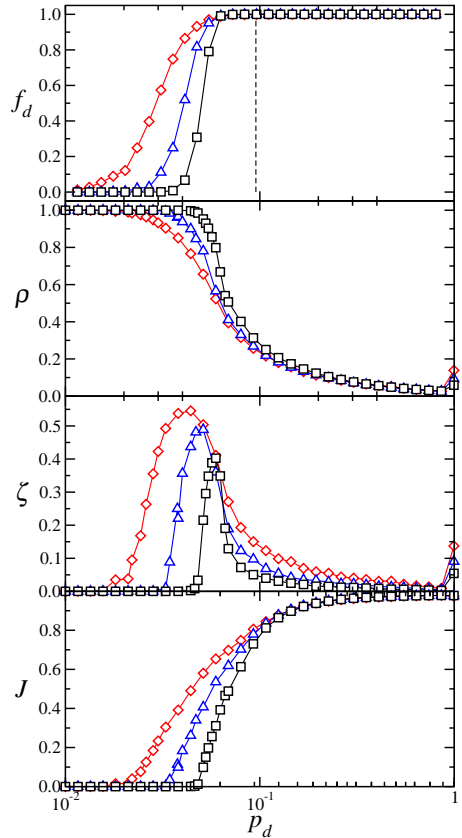}}
\end{center}
\vspace*{10 pt} \caption{Uppermost panel: Fraction $f_d$ of
realisations of the network of repulsive links for which the fully
synchronised state is unstable, as a function of the probability
$p_d$, over series of $10^3$ realisations. From left to right:
$N=50$ ($\diamond$), $100$ ($\vartriangle$) and $200$ ($\Box$). The
other parameters are $C=4$, $\kappa=10$, and $\omega=0.98$. The
vertical dotted line stands for the theoretical transition point,
$p_d^* \approx 9.54 \times 10^{-2}$, for an infinitely large
ensemble. The three lower panels show the Kuramoto order parameter
$\rho$, the Shinomoto-Kuramoto order parameter $\zeta$, and the
current $J$, for the same realisations.} \label{fig1}
\end{figure}

\begin{figure*}[t]\begin{center}
\resizebox{0.7\columnwidth}{!}{\includegraphics*{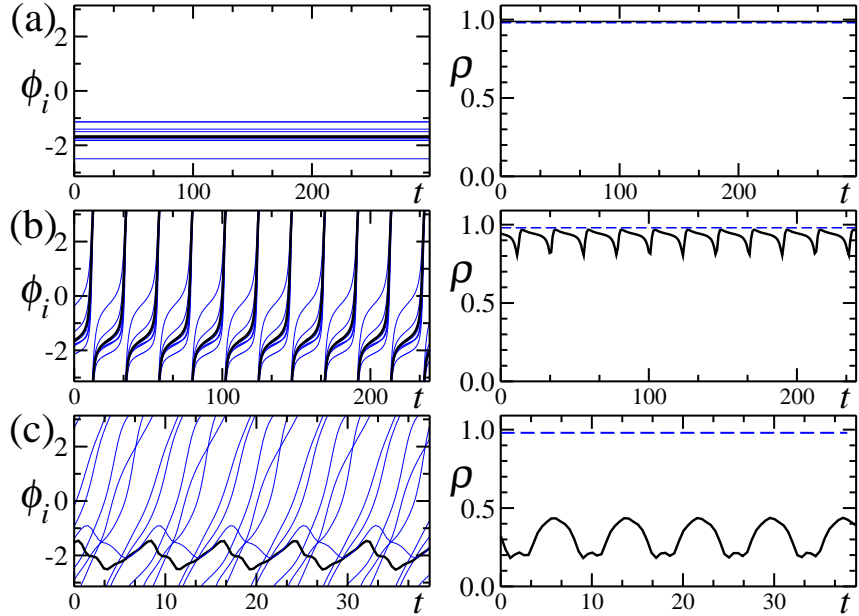}}
\end{center} 
\caption{Left column: Time evolution of the collective phase
$\Psi(t)$ (thick lines) and of the phases of ten representative
rotators (thin lines) in an ensemble of size $N=50$, for three
values of the fraction of repulsive links: (a) $p_d=0.015$, (b)
$0.03$, and (c) $0.1$. The other parameters are $C=4$, $\kappa=10$,
and $\omega=0.98$. Right column: Time evolution of $\rho(t)$, for
the same realisations. The horizontal dashed line stands for the
threshold of oscillatory behaviour, $\rho =\omega$ \cite{TSCT:2007}.
The middle panels correspond to the regime of global firing.}
\label{fig2}
\end{figure*}

Numerical results suggest that, for large $N$, there is a sudden
transition from $f_d=0$ to $1$ at a given value $p_d^*$ of the
probability $p_d$. An analytical estimate of $p_d^*$ can be obtained
from the assumption that the number of neighbours in the network of
repulsive links is the same for all rotators, i.e. that the number
of rotators in ${\cal N}_j$ is the same for all $j$. This
approximation improves as $N$ grows, because --as stated above-- the
relative dispersion in the number of rotators in ${\cal N}_j$
decreases as $1/\sqrt{N}$. It implies that the matrix  $\mathbb{J}$
can be written as
\begin{equation} \label{eq:matj2b}
 \mathbb{J} = - ( \sqrt{1-\omega^2} + C -C(\kappa+1)p_d)
\mathbb{I} - \frac{C (\kappa+1) }{ N}  \mathbb{M} .
\end{equation}
For $N\to\infty$, the coefficient in front of the matrix
$\mathbb{M}$ tends to zero, suggesting that its contribution to
$\mathbb{J}$ can be neglected in that limit. This intuitive argument
implies that the point at which the maximum eigenvalue becomes
positive for $N\to \infty$ is
\begin{equation}\label{eq:pd}
p_d^* = \frac{ C+\sqrt{1-\omega^2}}{C(1+\kappa)}.
\end{equation}
A more rigorous argument, confirming this critical value for
$p_d^*$, can be obtained from exact bounds for the eigenvalues of
the matrix $\mathbb{J}$ in Eq. (\ref{eq:matj}) coming from the
so-called semicircle law \cite{semic}, applied to matrices
$\mathbb{M}$ and $\mathbb{N}$ \cite{DHZ:2005}. The vertical dotted
lines in the uppermost panel of Fig. \ref{fig1} stands for this
critical value $p_d^*$.

Close inspection of the destabilisation of full synchronisation
reveals that this transition is triggered by the behaviour of the
rotator $j^*$ with the largest number of repulsive links --as also
observed to occur in ensembles of coupled phase oscillators
\cite{DHZ:2005}. For a given value of $\kappa$, as the fraction
$p_d$ of repulsive links grows, the fixed point $\phi_s$ first ceases to
be a stable state for $\phi_{j^*}$. An arbitrarily small deviation
in $\phi_{j^*}$ leads to a different equilibrium point, and full
synchronisation breaks down, even when other rotators may remain
mutually synchronised. For small sizes $N$, the value of $p_d$ at
which the rotator $j^*$ attains the number of negative interaction
weights which makes full synchronisation unstable depends sensibly
on the detailed structure of the network of repulsive links. This
explains the smooth transition in $f_d$ as $p_d$ grows. For large
$N$, on the other hand, all sites in the network of repulsive links
are statistically equivalent with respect to the distribution of
negative links. Consequently, the transition takes  always place at
the same value of $p_d$, irrespectively of the specific realisation
of the network. Accordingly, $f_d$ exhibits a sharp growth from $0$
to $1$, which should become discontinuous in the thermodynamic
limit. This semi-qualitative analysis sheds light on the role of the
homogeneity of the distribution of negative links in defining the
nature of the destabilisation transition of full synchronisation.
This effect of the network structure on the collective dynamics of
the system is further studied in the following.

\subsection{Numerical study of the unsynchronised regime}

In the  fully synchronised  state, the phases of all rotators are at
the stable fixed point $\phi_s$. The Kuramoto order parameter
reaches its maximum value $\rho=1$, whereas the absence of any kind
of collective motion implies that both the Shinomoto-Kuramoto
parameter $\zeta$ and the current $J$ vanish. As full
synchronisation breaks down, the parameter $\rho$ is expected to
decrease, and $\zeta$ and $J$ can adopt non-zero values, as was
shown in \cite{TSCT:2007}. In this section, we present numerical
results illustrating the behaviour of the order parameters as a
function of the fraction of repulsive links $p_d$, for ensembles of
different sizes,  with  $C=4$,  $\kappa=10$, and $\omega=0.98$.

Figure \ref{fig2} illustrates the dynamics of the rotator ensemble
for three values of the fraction of $p_d$. Results correspond to
numerical integration of Eq. (\ref{ard:phidotredux}) for a system of
$N=50$ rotators. In the left column, we show the time evolution of
the collective phase $\Psi (t)$  together with that of ten
representative rotators. In  panel (b), which corresponds to an
intermediate value of $p_d$, the rotators pulse in synchrony and,
consequently, the collective phase performs periodic rotations. We
identify this behaviour with the regime of global firing. In the
right column, we show the evolution of the parameter $\rho (t)$,
defined in Eq. (\ref{eq:rho}), for the same realisations. The degree
of synchronisation, measured by the time average of $\rho (t)$,
decreases as the fraction of repulsive interactions grows.

In the three lower panels of Fig. \ref{fig1} we plot the order
parameters $\rho$, $\zeta$, and $J$ as functions of the fraction of
repulsive links. As expected, the decay in $\rho$  is initially accompanied by
growth in both $\zeta$ and $J$. As $p_d$ increases, the current
keeps growing indicating that, on average, the firing frequency of
individual rotators is also increasing. The Shinomoto-Kuramoto
parameter $\zeta$ attains a maximum at an intermediate
value of $p_d$ and then decreases. This is an indication that global
synchronous firing occurs when full synchronisation is unstable and
for moderate values of $p_d$. Larger fractions of negative links,
however, make the dynamical coherence of the population decrease,
and global firing becomes less distinct.

These results --in particular, the existence of and intermediate
range of the probability $p_d$ where global firing is most
conspicuous-- are in qualitative agreement with the collective
behaviour of rotator ensembles where the disorder associated with
repulsive interactions is replaced by noise and/or ensemble
heterogeneity \cite{TSCT:2007}. However, a noticeable quantitative
difference resides in that, in the present case, the range
of global firing becomes narrower as the system grows in size. The
results of Fig. \ref{fig1} suggest that global firing may disappear
for sufficiently large ensembles. An explanation for this collapse
can be put forward recalling that the instability transition of full
synchronisation becomes sharper as $N$ grows (cf. the uppermost
panel of Fig. \ref{fig1}). As discussed at the end of section
\ref{sect:synchr}, for large systems, the number of repulsive links
is statistically the same for all rotators. At the transition, all
rotators abandon simultaneously the stable fixed point $\phi_s$ and,
consequently, the ensemble adopts a highly disorganised phase
configuration. In small systems, on the other hand, the transition
is gradual. The stable fixed point is first abandoned by those
rotators with a large number of repulsive links, which can be thus
entrained into global synchronous firing.

This explanation indicates that, if it were possible to design a
random network where the heterogeneity in the distribution of
repulsive links is maintained as the system grows in size, the
regime of global firing would persist for arbitrary large systems.
In section \ref{sect:hier} we introduce a procedure to construct
such a heterogeneous network, and numerically verify that the range
where global firing occurs does not collapse as $N$ grows. We also
propose an interpolation between that kind of heterogeneous networks and
Erd\H{o}s-R\'enyi networks, and analyse global firing on these
intermediate structures.

\section{Heterogeneous networks} \label{sect:hier}

In this section, we introduce a method to construct a class of
networks where the distribution of links per site preserves its
heterogeneity as the size $N$ increases. Specifically, the
dispersion in the number of links per site is, in these networks,
proportional to $N$ so that the relative width of the distribution
does not change with the size. According to the discussion at the
end of section \ref{sect:erdos}, the effects of heterogeneity
observed in small random networks of repulsive links, which
disappear because of self-averaging in the random distribution of
links, should persist in these new structures. In particular, we
expect to find global firing even in large ensembles.

\begin{figure}\begin{center}
\resizebox{.40\columnwidth}{!} {\includegraphics*{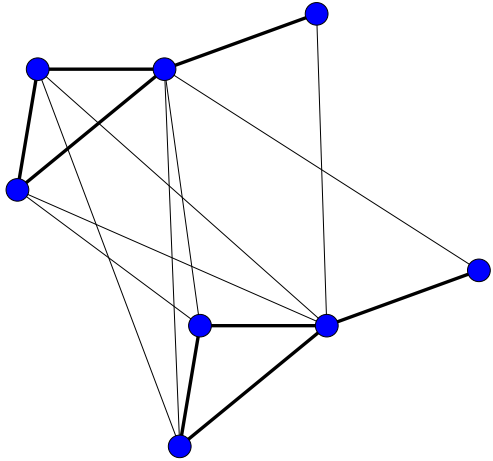}}
\end{center}
\vspace*{20 pt} \caption{A heterogeneous network constructed
starting by a network of $N_0=4$ sites, and consisting of two
replicas of the starting network. Bold lines stand for links inside
each replica, and thin lines correspond to links connecting
different replicas.} \label{fig3}
\end{figure}

\subsection{Construction of a heterogeneous network}
\label{sect:4.1}

To build up our heterogeneous network of repulsive links, we start
by taking a random network of size $N_0$ and wiring probability
$p_d$. The average number of links per site is  $\bar k_0\sim
p_dN_0$, while its mean square dispersion is $\sigma_0 \sim
\sqrt {p_d N_0}$. Let ${\cal N}_j$ be the set of the $k_j$
neighbours of site $j$. We now consider $M$ identical replicas of
the starting network, which we use to build up a network of size
$N=MN_0$. In each one of these replicas, say in replica $m$, we
identify the site $j_m$ which is homologous to the site $j$ of the
starting network. The $M$ replicas are linked by connecting each
site $j_m$ with all the neighbours of all its homologous sites in
all the replicas. In other words, the site $j_m$ becomes linked to
the $k_j M$ members of the sets ${\cal N}_{j_1}$, ${\cal N}_{j_2}$,
..., ${\cal N}_{j_M}$. Figure \ref{fig3} illustrates the resulting
network in a simple case. By construction, the number of neighbours
of each site in the final structure equals $M$ times the number
neighbours of the homologous site in the starting network.
Therefore, the average number of links per site and the mean square
dispersion are, respectively,
\begin{equation}
\bar k = M p_d N_0 =  M \bar k_0=p_d N ,
\end{equation}
and
\begin{equation}
\sigma = M \sigma_0=N \sigma_0/N_0.
\end{equation}
As advanced  above, for fixed $N_0$, both the mean value and the
dispersion in the distribution of links of the resulting network
grow linearly with the number of sites $N$. The relative dispersion
is thus independent of the network size.

Let $\mathbb{M}_0$ and $\mathbb{M}$ be the adjacency matrices of the
starting network and the resulting network, respectively. It is
possible to show that, if $\lambda_0$ is an eigenvalue
$\mathbb{M}_0$, then $\lambda=M\lambda_0=N\lambda_0/N_0$ is an
eigenvalue of $\mathbb{M}$, with multiplicity $N_0$. The remaining
eigenvalues are equal to zero. A similar relation holds for the
matrix $\mathbb{N}$ of Eq. (\ref{eq:matj}). This result differs from
the prediction of the semicircle law, which stands for purely random
matrices and states that $\lambda \sim \sqrt{N}$  \cite{semic}.
Since, in Eq. (\ref{eq:matj}), both $\mathbb{M}$ and $\mathbb{N}$
are divided by the network size $N$, the maximum eigenvalue of
$\mathbb{J}$ --which is in turn determined by the maximum (positive)
eigenvalue of $\mathbb{M}$-- does not depend on $N$. Thus, for fixed
$N_0$, the critical point at which full synchronisation becomes
unstable is independent of the size.

Clearly, the network resulting from our construction is not fully
random. In fact, because of the interconnection of all the
homologous sites, it can be shown to have a large clustering
coefficient \cite{AB:2002}. To interpolate between this structure
and a fully random network, we introduce a mechanism of link
rewiring, in the spirit of small-world networks \cite{WS:1998}.
Starting from the heterogeneous network constructed above, we visit
each site and, with probability $p_f$, rewire each of its links to a
randomly chosen site all over the system. For $p_f=1$, a fully
random network is obtained. In the following, we study the
collective behaviour of an ensemble of globally coupled rotators whose
network of repulsive interactions is build as described here, in
terms of the parameters $p_d$ and $p_f$.

\begin{figure}\begin{center}
\resizebox{.37\columnwidth}{!}{\includegraphics*{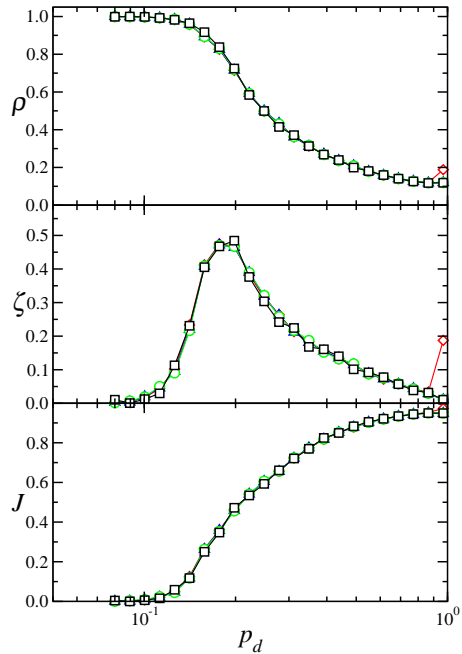}}
\end{center} \caption{The order parameters as functions of the
fraction of repulsive links $p_d$, in the absence of rewiring,
$p_f=0$. Different curves, which are however practically coincident,
correspond to sizes $N=100$, $200$, $400$, and $1000$. In all cases,
the network of negative links is constructed from a starting network
of $N_0=20$ sites. The other parameters are $C=4$, $\kappa = 2$, and
$\omega = 0.98$.} \label{fig4}
\end{figure}
\begin{figure}\begin{center}

\resizebox{.37\columnwidth}{!}{\includegraphics*{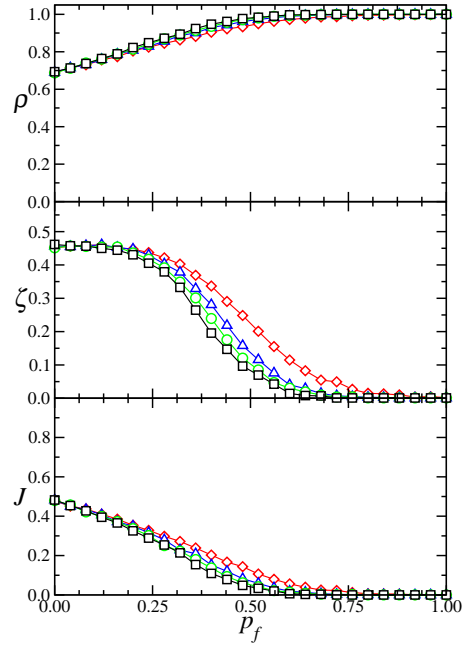}}
\end{center} \caption{The order parameters as  functions of the
rewiring probability $p_f$, for a fixed fraction of repulsive links,
$p_d=0.2$. Different curves correspond to sizes $N=100$
($\diamond$), $200$ ($\vartriangle$), $400$ ($\circ$), and $1000$
($\Box$). In all cases, the network of negative links is constructed
from a starting network of $N_0=20$ sites. The other parameters are
$C=4$, $\kappa = 2$, and $\omega = 0.98$.} \label{fig5}
\end{figure}

\subsection{Numerical results}

In our numerical analysis, the networks of repulsive links
introduced in section \ref{sect:4.1} are built starting from random
networks of $N_0=20$ sites, and their final sizes are $N=100$,
$200$, $400$, and $1000$. Also, we fix $C=4$, $\kappa = 2$, and
$\omega = 0.98$.

Figure \ref{fig4} shows results for the order parameters $\rho$,
$\zeta$, and $J$ as functions of the fraction of repulsive links
$p_d$, in the absence of rewiring, $p_f=0$. Different curves
correspond to different sizes $N$. We see that, as expected, the
order parameters are essentially independent of the system size. In
particular, the regime of global firing --signalled, for intermediate
values of $p_d$, by the maximum in the Shinomoto-Kuramoto
parameter-- clearly persists for large ensembles.

Figure \ref{fig5} shows the order parameters as functions of the
rewiring probability $p_f$, for a fixed fraction of negative links,
$p_d=0.2$ which corresponds, approximately, to the maximum common
firing activity in the case $p_f=0$, see figure \ref{fig4}. The
Shinomoto-Kuramoto parameter $\zeta$ is practically independent of
$N$ for small and large $p_f$, but exhibits a substantial dependence
on the size for intermediate wiring probabilities. As may be
expected from our results on Erd\H{o}s-R\'enyi networks, in this
intermediate zone, $\zeta$ is larger for smaller ensembles.

Finally, Fig. \ref{fig6} summarises the dependence of the order
parameters on the probabilities $p_d$ and $p_f$ for an ensemble of
$N=400$ rotators. The graphs represent numerical results averaged
over series of $100$ realisation for each parameter set.

\begin{figure}
\begin{center}
\resizebox{.37\columnwidth}{!}{\includegraphics*{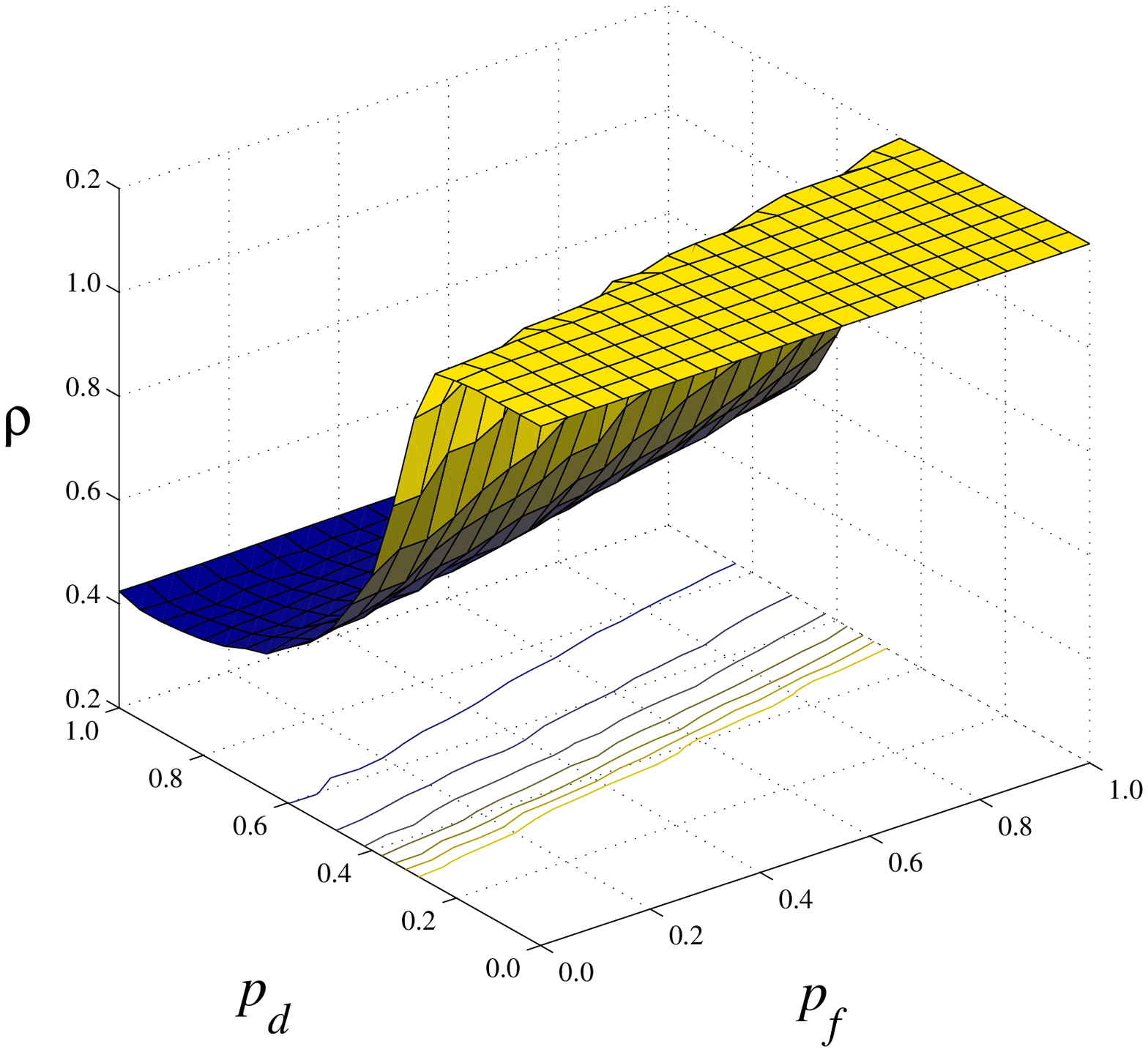}}
\resizebox{.37\columnwidth}{!}{\includegraphics*{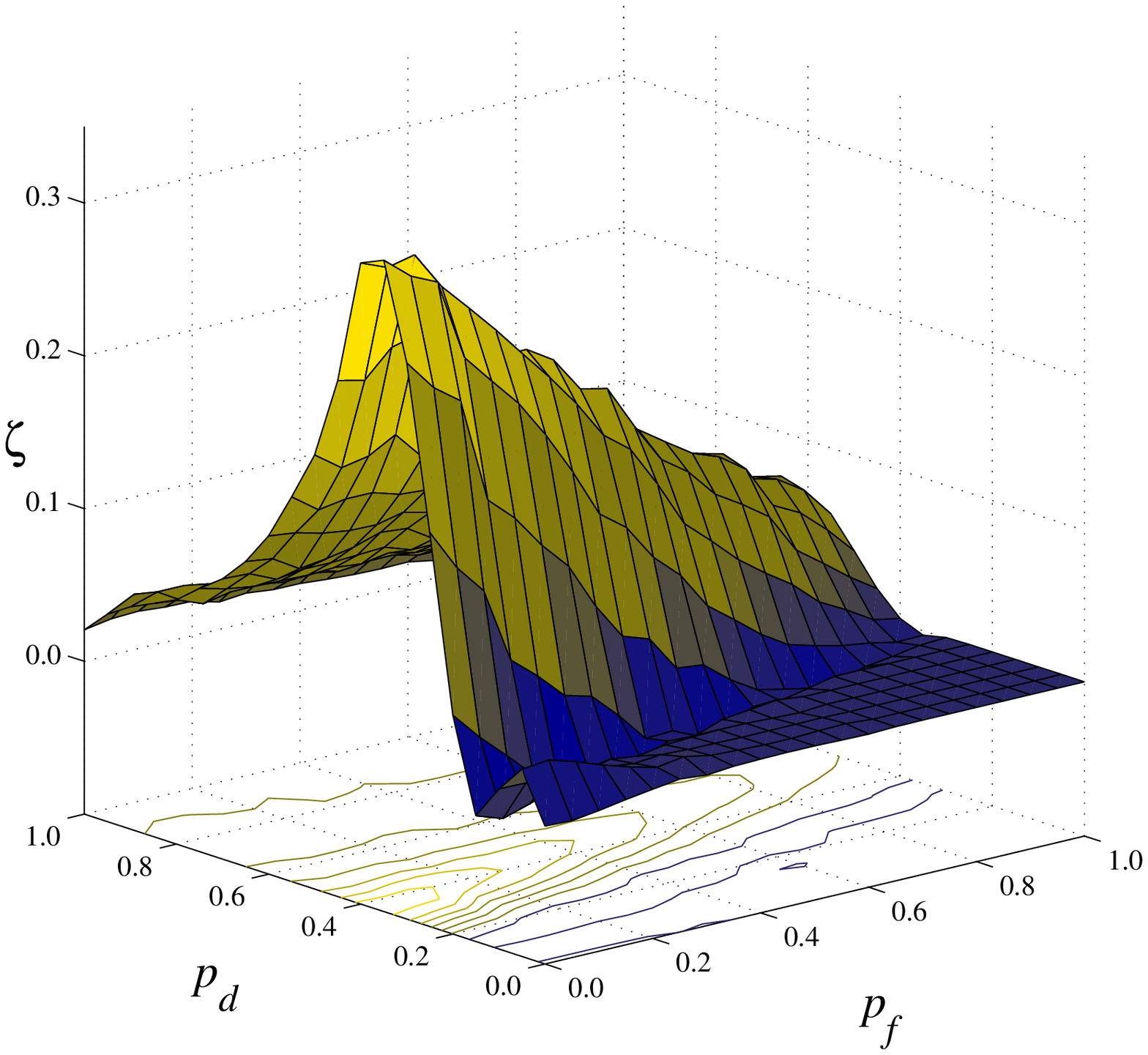}}
\resizebox{.37\columnwidth}{!}{\includegraphics*{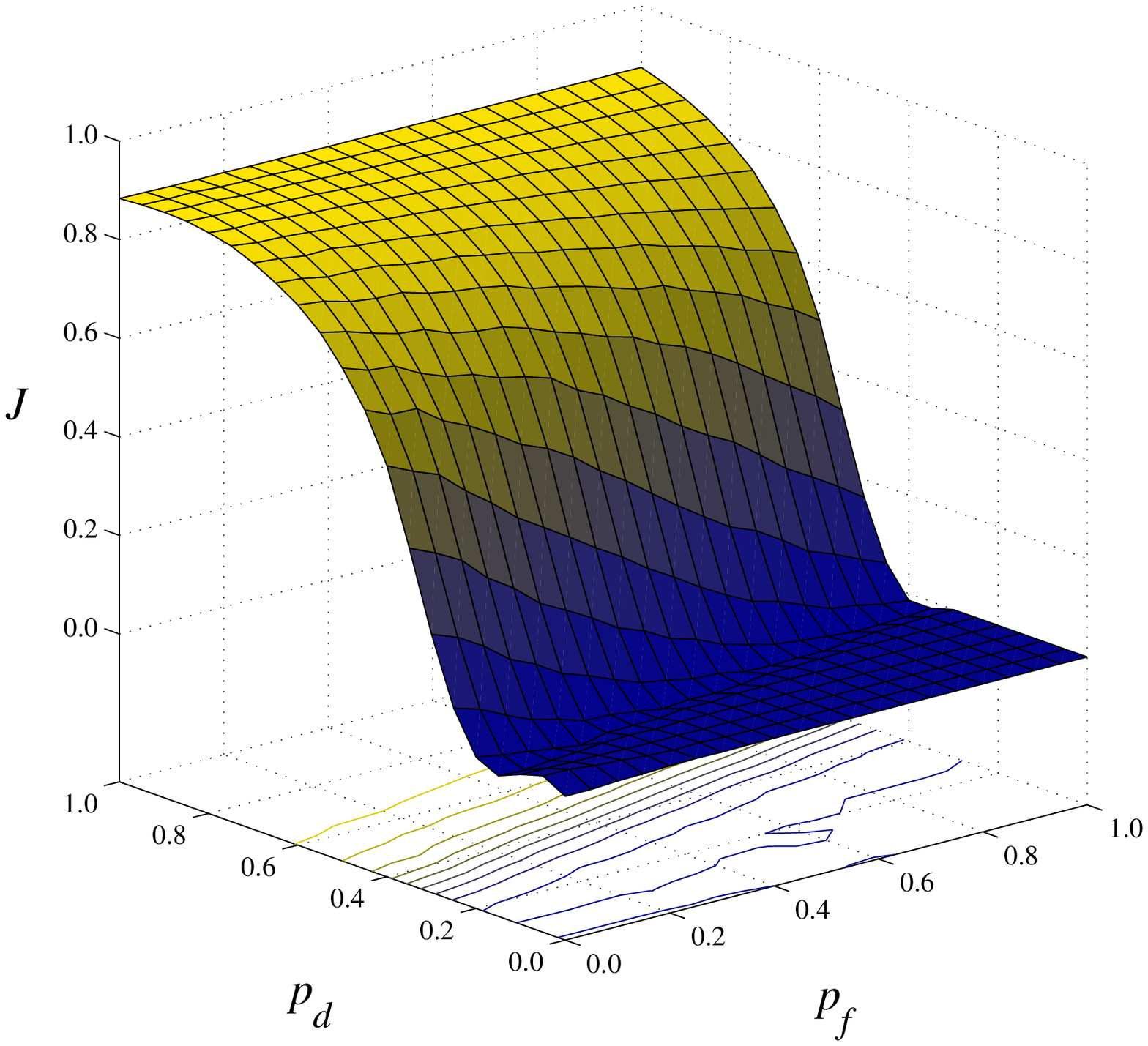}}
\end{center}
\caption{The order parameters for an ensemble of $N=400$ rotators
($N_0=20$) as functions of the fraction of negative links $p_d$ and
the rewiring probability $p_f$. The other parameters are $C=4$,
$\kappa=2$, and  $\omega=0.98$.} \label{fig6}
\end{figure}

\section{Conclusion}
\label{sect:conc} In this paper, we have shown that --like noise and
ensemble heterogeneity \cite{TSCT:2007}-- structural disorder in a
network of interacting excitable rotators can induce global firing
of the ensemble. This form of collective dynamics is here a
consequence of the presence of a fraction of repulsive couplings,
randomly scattered all over the interaction network. These repulsive
interactions destabilise the state of full synchronisation when
their number and intensity are large enough. Just beyond this
desynchronisation transition, for intermediate values of the
fraction of repulsive couplings, the ensemble is entrained in a
regime where many rotators are coherently excited, and global firing
is triggered. While the results of this paper support the general
mechanism for global firing put forward in \cite{TSCT:2007}, it is
important to stress that the particular mechanism at work in the
system studied here is quite different. In \cite{TSCT:2007} the
rotators showed heterogeneity in the form of dispersion in the
natural frequencies $\omega_i$. As the dispersion increased a
fraction of the rotators had natural frequencies in the oscillatory
range $\omega_i>1$. Those rotators would spontaneously fire if they
were isolated. When coupled, they pull the other rotators into an
state of collective firing. In the system studied in this paper, on
the other hand, the mechanism of destabilisation is produced by the
effect that negative couplings have on individual rotators, but
these would never fire if isolated. In both cases, however, it is
the diversity, either in the natural frequencies or in the number of
negative links, that produces the global effect. In fact, when the
pattern of repulsive couplings forms an Erd\H{o}s-R\'enyi random
network the regime of global firing is progressively
suppressed as the size of the ensemble grows. This effect can be
ascribed to that, as the network becomes larger, there is
self-averaging in its degree distribution, so that the number of
neighbours per site becomes increasingly homogeneous. This
homogeneity sharpens the desynchronisation transition and, at the
same time, causes the collapse of the zone where global firing is
possible for smaller systems. To avoid this effect of self-averaging
--which clearly illustrates how increasing homogeneity can inhibit
certain forms of coherent behaviour-- we have introduced an
algorithm to construct partially random networks whose degree
distribution maintains its relative dispersion as the network grows.
In this case, in fact, the order parameters which characterise the
macroscopic behaviour of the rotator ensemble become essentially
independent of the size. In particular, the regime of global firing
persists for large systems.

The present results extend previous work on the effects of diversity
of various origins over the collective dynamics of complex systems.
These forms of diversity include now heterogeneous interaction
patterns, with positive (attractive) and negative (repulsive)
couplings. The extension is thus relevant to such systems as neural
networks, where interactions of different signs are present in the
form of activator and inhibitory synapses \cite{DA:2001}.
\vspace{1.0cm}

We acknowledge financial support from the MEC (Spain) and FEDER (EU)
through projects FIS2006-09966, FIS2007-60327, the EU NoE BioSim
LSHB-CT-2004-005137, and from CONICET and ANPCyT (Argentina) through
projects PIP5114 and PICT04-943. CJT acknowledges financial support from SBF
(Swiss Confederation)
through research project C05.0148 (Physics of Risk).

\end{document}